\documentclass[apj]{emulateapj}
\usepackage{ifpdf}
\pdfoutput=1
\usepackage{graphicx}
\usepackage{natbib}
\bibpunct{(}{)}{;}{a}{}{,}

\newcommand{\ee}[1]{\mbox{${} \times 10^{#1}$}}

\newcommand{\degree}{\mbox{$^{\circ}$}}

\newcommand{\kms}{\mbox{km s$^{-1}$}}

\newcommand{\lsun}{\mbox{L$_\odot$}}
\newcommand{\msun}{\mbox{M$_\odot$}}
\newcommand{\rsun}{\mbox{R$_\odot$}}

\newcommand{\tr}{\mbox{$T_R$}}
\newcommand{\tk}{\mbox{$T_K$}}

\newcommand{\rinf}{\mbox{$r_{{\rm inf}}$}} 
\newcommand{\cseff}{\mbox{$c_{{\rm s,eff}}$}} 

\newcommand{\form}{H$_2$CO}

\newcommand{\ammonia}{\mbox{{\rm NH}$_3$}}

\newcommand{\cooo}{C$^{18}$O}

\newcommand{\hcop}{HCO$^+$}

\newcommand{\hccn}{\mbox{H$^{13}$CN}}

\newcommand{\jj}[2]{\mbox{$J = #1\rightarrow#2$}}

\newcommand{\minf}{\mbox{$\dot M_{\rm inf}$}}
\newcommand{\mstar}{\mbox{$M_{\star}$}}
\newcommand{\lstar}{\mbox{$L_{\star}$}}
\newcommand{\rstar}{\mbox{$R_{\star}$}}

\newcommand{\msunyr}{\mbox{M$_\odot$ yr$^{-1}$}}

\newcommand{\mdotacc}{\mbox{$\dot M_{\rm acc}$}}
\newcommand{\mdotwind}{\mbox{$\dot M_{\rm w}$}}
\newcommand{\facc}{\mbox{$f_{\rm acc}$}}

\shorttitle {B335 Infall from ALMA Observations}
\shortauthors{Evans et al.}

\begin{document}

\title{Detection of Infall in the Protostar B335 with ALMA}
\author{Neal J. Evans II\altaffilmark{1}, 
James Di Francesco\altaffilmark{2},
Jeong-Eun Lee\altaffilmark{3},
Jes K. J{\o}rgensen\altaffilmark{4},
Minho Choi\altaffilmark{5},
Philip C. Myers\altaffilmark{6},
Diego Mardones\altaffilmark{7}
}
\altaffiltext{1}{Department of Astronomy, The University of Texas at Austin,
2515 Speedway, Stop C1400, Austin, Texas 78712-1205, U.S.A.; 
nje@astro.as.utexas.edu}
\altaffiltext{2}{ National Research Council of Canada, Herzberg Institute of 
Astrophysics, 5071 West Saanich Road, Victoria, BC V9E 2E7, Canada}
\altaffiltext{3}{School of Space Research, Kyung Hee University, Yongin-shi, Kyungki-do 449-701, Korea}
\altaffiltext{4}{Centre for Star and Planet Formation, Niels Bohr Institute and Natural History Museum of Denmark, {\O}ster Voldgade 5--7, 1350 Copenhagen K., Denmark}
\altaffiltext{5}{ Korea Astronomy and Space Science Institute, 776 Daedeokdaero, Daejeon 305-348, Korea }
\altaffiltext{6}{Harvard-Smithsonian Center for Astrophysics, 60
  Garden Street, Cambridge, MA 02138, U.S.A.}
\altaffiltext{7}{Departamento de Astronom{\'i}a, Universidad de Chile, Casilla 36-D, Santiago, Chile}
\email{nje@astro.as.utexas.edu}

\begin{abstract}
Observations of the isolated globule B335 with ALMA have yielded
absorption features against the continuum that are redshifted from the
systemic velocity in both HCN and \hcop\ lines. These features provide
unambiguous evidence for infall toward a central luminosity source.
Previously developed models of inside-out collapse can match
the observed line profiles of HCN and \hcop\ averaged over the central
50 AU. At the new distance of 100 pc, the inferred infall radius is
0.012 pc, the mass infall rate is 3\ee{-6} \msunyr,
the age is 5\ee4 years, and the accumulated mass in the central zone
is 0.15 \msun, most of which must be in the star or in parts of a disk that
are opaque at 0.8 mm. The continuum detection indicates an optically
thin mass (gas and dust) 
of only 7.5\ee{-4} \msun\ in the central region, consistent
with only a very small disk mass.
\end{abstract}


\section{Introduction}

The idea that stars form from the infall of material from a dense
core in a molecular cloud is enshrined in all cartoons of star formation.
By {\bf infall}, we mean the inward motion of matter toward
a central forming star, driven by self-gravity. 
While such a picture is attractive theoretically, 
hard observational evidence of such infall has proven to be elusive.

Most evidence in support of infall has been indirect and depends on
comparison of observations of molecular line emission profiles
to predictions of calculations that involve complexities of chemistry
and excitation, along with assumptions about density, temperature,
and velocity fields \citep{2000prpl.conf..217M}.

It was realized early 
\citep{1977ApJ...214L..73L}
that red-shifted absorption against a central
continuum source would provide an unambiguous indicator of matter in
front (absorption) moving inward (redshifted) toward that source.  If the
envelope emits
at velocities centered on the mean envelope velocity in the spectral line
chosen, the result is like an inverse P-Cygni profile (IPC), with emission 
on the blue side of the central velocity and absorption on the red side. 
Because the opacity of dust grains increases rapidly with frequency at
millimeter and submillimeter wavelengths, absorption will be easier
to detect at higher frequencies.
\citet{2012A&A...542L..15W}
have taken advantage of this fact to find red-shifted absorption
features in \ammonia\ rotational transitions at THz frequencies
toward a number of massive star-forming clumps, strongly indicative of 
inflow of material within the clumps.
By {\bf inflow}, we mean general inward motion toward a forming cluster
of stars, rather than infall toward a single star.
\citet{2013A&A...558A.126M} 
also used THz frequencies, together
with the fact that gas-phase water reaches substantial abundances
only in the warm inner regions, 
to identify IPCs in several regions of lower mass star formation.
One drawback of using water is that one must remove the strong emission
from the outflow. That is not a problem for observations of starless
cores, and
\citet{2015MNRAS.446.3731K}
have used IPCs in water lines to detect subsonic contraction motions
in L1544, consistent with those predicted for a quasi-equilibrium
Bonner-Ebert sphere.

These studies have used {\it SOFIA} and {\it Herschel}
observations and thus have spatial resolutions of 16\arcsec\ to 30\arcsec.
Another approach is to improve the spatial resolution using interferometers
\citep{1999ApJS..122..519C}.
\citet{2001ApJ...562..770D}
exploited this ability to see 
inverse P-Cygni profiles toward NGC1333 IRAS4.
In that rather complex region, there were suggestions that
the absorption arose from resolving out emission 
in an extended foreground region with a slight
velocity offset, rather than in an infalling envelope
\citep{2004ApJ...617.1157C}.
Since this source was one of the few with sufficiently strong 
continuum emission to observe the effect with earlier interferometers, 
this technique was largely 
put on the shelf, ``waiting for ALMA."

We can now  use ALMA's high resolution and sensitivity 
to observe molecular lines in {\it absorption}\/ against the 
continuum emission from the disk or inner envelope.   Once
detected, absorption observed in a number of lines can be used to study
the velocity field of infall and the rate of central mass accumulation.  More
generally, these data can provide an independent constraint on infall
models based solely on emission line profiles. Another advantage
of very high spatial resolution is that the outflow region may be excluded,
allowing study of the higher infall velocities closer to the
central source.

The ideal source to test infall models is B335, an isolated, roundish,
dark globule (see images in 
\citealt{2007A&A...475..281G}). 
It was the first source
for which kinematic evidence of infall was claimed.
\citet{1993ApJ...404..232Z}
found a spectral signature in several lines of CS and \form\ toward B335.
The spectral signature, often called an ``asymmetrically blue profile"
because the blue peak (from the back of an infalling envelope) is stronger
than the red peak (from the front of an infalling envelope), requires
particular combinations of density, temperature, abundance, and velocity
fields, which may not be realized in a given source
(see, e.g., \citealt{1995ApJ...448..742C}).
Later, follow-up observations with higher resolution raised doubts
about the infall interpretation because of contributions to emission from
outflowing gas \citep{2000ApJ...544L..69W}.
The recently revised distance to B335 of $105 \pm 15$ pc 
\citep{2009A&A...498..455O}
makes it even more attractive for a close-up look. 
We take a distance of 100 pc, 0.25 times the distance assumed in the
past, for convenience. As a result, the observed luminosity is 0.72
\lsun.
Finally, the outflow \citep{1988ApJ...327L..69H}
is nearly perpendicular to the line of sight, with
an inclination angle of 87\degree\ \citep{2008ApJ...687..389S}.

\citet{2005ApJ...626..919E}
developed detailed models of line spectra toward B335 available at that time,
using inside-out 
\citep{1977ApJ...214..488S}
infall models and chemical modeling.
Recent studies of B335 
\citep{2011ApJ...742...57Y, 2010ApJ...710.1786Y}
have confirmed the results of earlier models that found B335 to be a very
slowly rotating core in the inner regions. 
A detailed model of infall was developed by 
\citet{2013ApJ...765...85K}.
Most recently,
\citet{2015ApJ...799..193Y}
fitted data from the SMA to a model of essentially pure infall, with
an upper limit on specific angular momentum in the inner regions of
$j < 5\ee{-5}$ \kms pc, and they predicted that a rotationally supported
disk would  have a radius no larger than 5 AU. 
\citet{2015arXiv150904675Y} have 
confirmed the slow rotation with ALMA data on \cooo\ and SO and set an
upper limit of 10 AU on the  radius of any Keplerian disk.

\section{Observations}

Observations of B335 were obtained as part of project 2012.1.00346.S (PI. N.
Evans) on 27 April 2014 (UT) in the C32-3 configuration of ALMA Early Science.  
The observations used 35 12-m antennas 
and the Band 7 receivers.  The weather conditions during the observations were 
reasonable, with roughly stable values of precipitable water vapor ranging 
between 1.0 mm and 1.1 mm.  Data of B335 were obtained over a baseline range 
of 18.69 m to 527.57 m. Half the flux of a Gaussian with FWHM size of 
4\farcs1 would be recovered with this minimum baseline (eq. A8 of 
\citealt{1994ApJ...427..898W}).

The ALMA Correlator was configured to have four spectral windows (spws), each 
with 3840 channels 61.035 kHz ($\sim 0.05$ \kms) 
wide and a total useable bandwidth of 234.375 MHz.  
(The spectral resolution is twice the channel width.)  The local
oscillator was tuned 
appropriately so that HCN 4-3, HCO$^{+}$ 4-3, CS 7-6, and H$^{13}$CN 4-3 could 
be observed in spws 0, 1, 2, and 3, respectively.  The rest frequencies, 
velocity resolutions, upper state energies above ground (in kelvin), and 
Einstein $A_{ul}$ values are given in Table \ref{obstab}.

The execution block lasted for 97 minutes total, and consisted of a $\sim$10
minute integration on J1924-2914 to calibrate the bandpass shape of each spw, 
then a $\sim$2.5 minute integration on J1751+096 to calibrate amplitudes.  
Observations of B335 were obtained with $\sim$7 minute integrations and were
interleaved with $\sim$1.5 minute integrations of J1955+1358 to calibrate 
phases.  B335 was observed for a total of 48.35 minutes.  The data were 
initially calibrated and imaged by staff at the North American ALMA Science 
Center (NAASC), using standard procedures in the Common Astronomy Software 
Applications (CASA) package.  Passing the QA2 quality assurance step, the data 
were delivered to the project team for further calibration and imaging, again 
using CASA.

Preliminary imaging of the NAASC-calibrated data of B335 revealed obvious 
detections of the continuum source in every channel.  A smoothed version of
the data was self-calibrated using line-free channels to create a continuum 
dataset.  The data were self-calibrated in phase first at 600 sec intervals 
followed by another self-calibration in phase at 60 sec intervals.  The data 
were then self-calibrated in amplitude over 3600 sec intervals.  All three
self-calibrations were then applied to the unsmoothed dataset.  A continuum
image was then produced by inverting the line-free channels into a field with
512 $\times$ 512 cells of $0\farcs1 \times 0\farcs1$
size each, using multi-frequency synthesis to populate the ${uv}$-plane.  The
dirty image was CLEANed using the basic Clark algorithm down to a threshold of 
0.5 mJy, roughly 2 $\sigma$.  The resolution of the CLEAN continuum image was 
$0\farcs486 \times 0\farcs439$, P.A. = $-76.9^{\circ}$.
The line data cubes were then produced by inverting each spectral window with 
61.035 kHz channels, respectively, with each channel having the same cell and 
field sizes as the continuum image.  Each channel was also CLEANed using the 
Clark algorithm down to a threshold of 60 mJy per channel, again $\sim$2 
$\sigma$.  The angular resolutions of the CLEAN datacubes were similar to that
of the continuum image.  
The continuum and line data were corrected for primary 
beam attenuation. Continuum subtraction in the uv plane can cause
artifacts in the presence of absorption features, so we removed
the continuum in the image plane.
Using MIRIAD 
\citep{1995ASPC...77..433S}, a continuum 
signal determined by fitting a polynomial to line-free channels was subtracted 
from the line data in the image plane.  The polynomials were first-order for 
the HCO$^{+}$, HCN, and CS cubes, but zeroth-order for the H$^{13}$CN cube due
to an abundance of lines in that spectrum.  The data were then written out as 
FITS files for further analysis.

\section{Results}\label{results}

\begin{figure}
\center
\includegraphics[scale=0.7, angle=0]{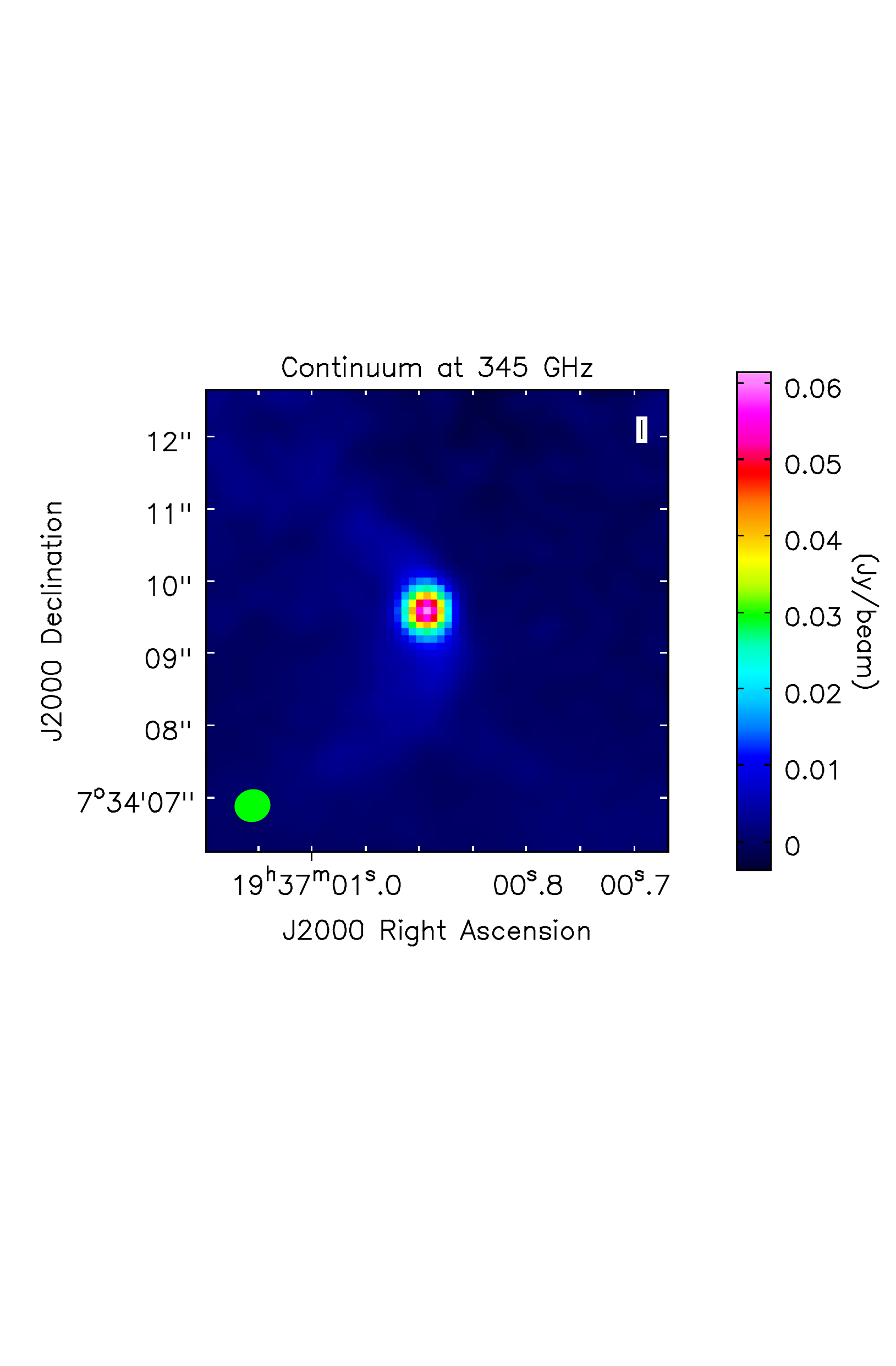}
\caption{
Continuum image of B335 at 345 GHz. The beam is shown in the lower
left corner as a green ellipse.
}
\label{contfig}
\end{figure}

The ALMA data show a very compact continuum source at 350 GHz
(Fig. \ref{contfig}).
A two-dimensional Gaussian fit yields a source centered at
19:37:00.894, $+$07:34:09.59 (J2000), with a deconvolved size (FWHM) of
$0\farcs45$ by $0\farcs25$ at position angle of $10\degree$, nearly
orthogonal to the beam and to the outflow axis. The 
total flux density is $91\pm6$  mJy.
There is also a faint, arc-like structure curving to the east, but appearing
to wrap around the continuum source.
\citet{2007ApJ...659..479J} found a continuum source at 0.8 mm at
19:37:00.91, $+$07:34:09.6 (J2000), with a size of $1\farcs6$, and
a total flux density of 350 mJy. If this source had been uniform over
their source size, we would have observed
34 mJy into our $0\farcs5$ beam, whereas we actually observed 91 mJy, 
indicative of a very compact structure.
At 100 pc, the structure would be 45 AU  by 25 AU.
This structure is most plausibly the inner part of the infalling envelope.
\citet{2015arXiv150904675Y} provide more detailed analysis of 
continuum data at similar resolution, obtained at lower frequencies,
and they also identify similar structures with comparable mass at this
scale.
No other compact source is present in the field down to a limit of 6 mJy
($3 \sigma$).

\begin{figure}
\center
\includegraphics[scale=1.00, angle=0, trim = 200 150 200 100]{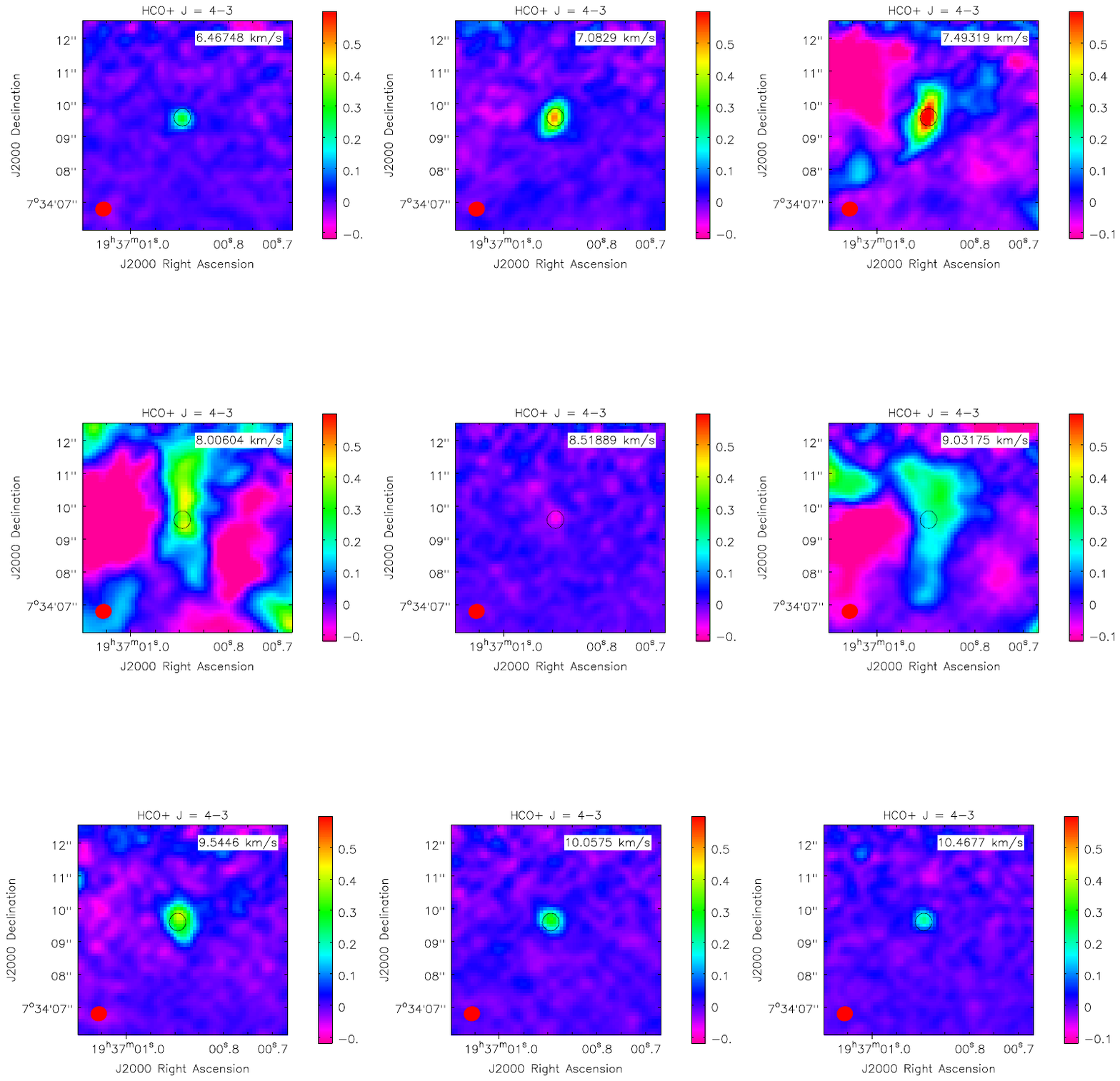}
\caption{
Channel maps of HCO$^+$ toward B335, spaced by 0.5 km s$^{-1}$,
after Hanning smoothing to an effective resolution of 0.2 \kms.
The intensity range runs from $-0.12$ Jy/beam to 0.6 Jy/beam
and scaling power cycle was set to $-0.2$ to display negative
intensity as purple. Rainbow 1 color scheme in CASA was used.
The red ellipse at lower left is the beam and the black ellipse near
center is the half-power contour for the continuum emission.
}
\label{HCOPchmap}
\end{figure}

\begin{figure}
\center
\includegraphics[scale=1.00, angle=0, trim = 200 150 200 100]{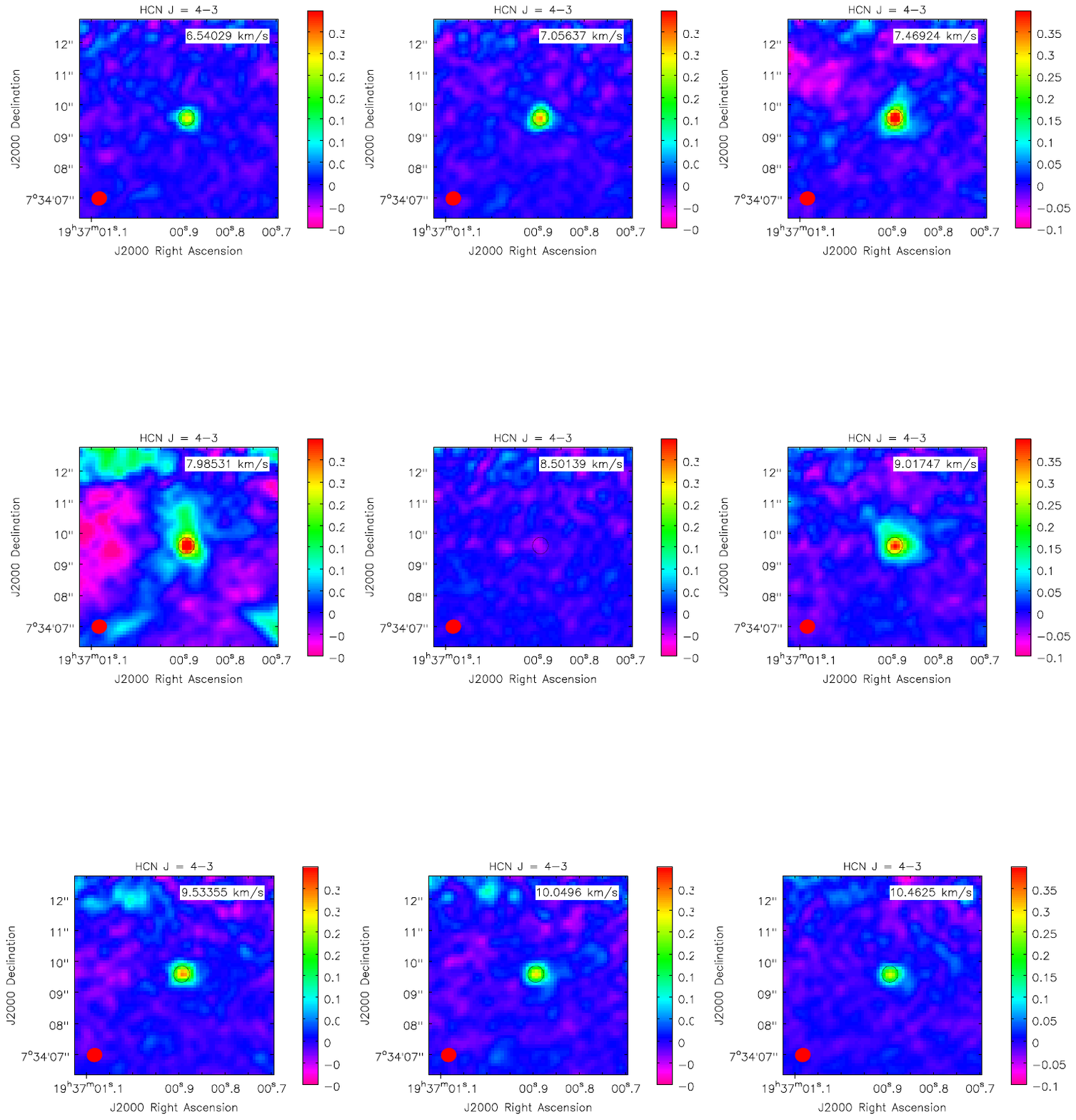}
\caption{
Channel maps of HCN toward B335, spaced by 0.5 km s$^{-1}$,
after Hanning smoothing to an effective resolution of 0.2 \kms.
The intensity range  runs from $-0.12$ Jy/beam to 0.6 Jy/beam
and scaling power cycle was set to $-0.2$ to display negative
intensity as purple. Rainbow 1 color scheme in CASA was used.
The red ellipse at lower left is the beam and the black ellipse near
center is the half-power contour for the continuum emission.
}
\label{chmaphcn}
\end{figure}

The spectral data show very compact (nearly point-like) emission centered
on the continuum peak, except near the systemic velocity of $8.30\pm 0.05$ \kms\
\citep{2005ApJ...626..919E}, where arc-shaped structures can be seen
in the CS line, more faintly in the \hcop\ line, and very faintly in
the HCN line. These structures are plausibly the edges of outflow
cavities. Channel maps were created after an additional 
Hanning smoothing to an effective resolution of 0.2 \kms.
The channel maps, spaced by about 0.5 \kms, are shown in Figure \ref{HCOPchmap}
for \hcop\ and Figure \ref{chmaphcn} for HCN. 
The \hcop\ emission in the line wings, and absorption at 8.5 \kms,
are both compact and centered on the continuum source (half-power emission
ellipse shown as a black contour), with no evidence of emission from the
outflow. At velocities near the line peaks, the emission is extended,
roughly north-south, and  separated emission, most likely from the cavity
walls of the outflow, can be seen about 3\farcs5 to the northeast and
southwest. False absorption can be seen inside
the outflow cavities, due to resolved out emission from the cavity walls,
at the line peak velocities, but {\bf not} at the velocity of the
absorption feature seen toward the continuum source. 
The channel maps for HCN (Figure \ref{chmaphcn})
are similar to those for \hcop, except that the emission from the
cavity walls is apparent only for the channel near the blue peak at
8.0 \kms, and the absorption at 8.5 \kms\ is less significant.

\begin{figure}
\center
\includegraphics[scale=0.7, angle=0]{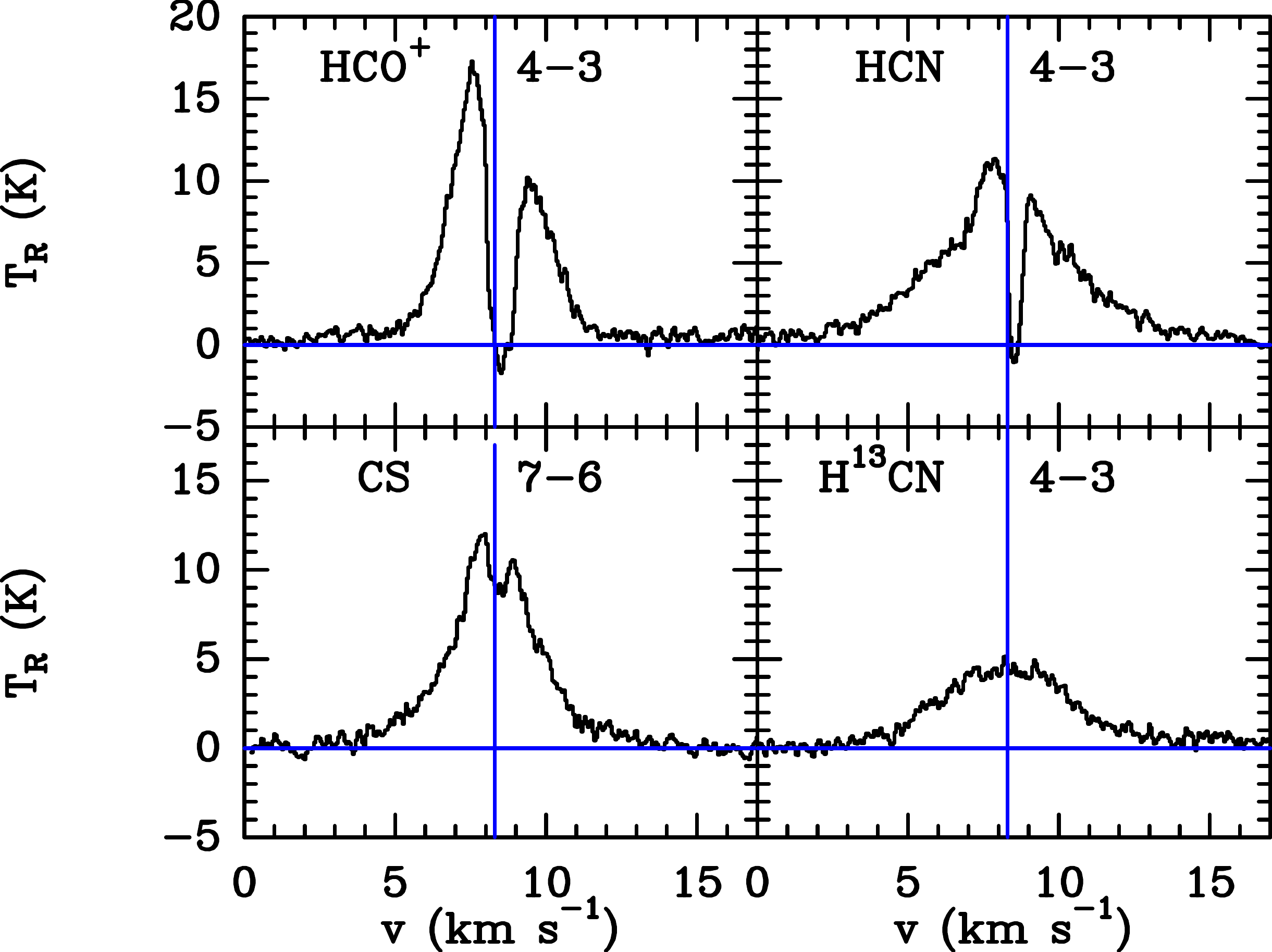}
\caption{
Spectra toward the continuum peak, after continuum subtraction,
and scaling to the $T_R$ scale. 
The blue horizontal line is at zero and the blue vertical line indicates
the systemic velocity of $8.30$ \kms\ found by
\citet{2005ApJ...626..919E}.
}
\label{almatr}
\end{figure}

The spectra for all four lines, extracted from a 0\farcs5 by 
0\farcs5 box centered on the 
continuum source, are shown in Figure \ref{almatr}. 
Except possibly for CS, contamination by emission from the outflow
cavity is negligible in this box.
The data were converted to a \tr\ scale using conversions from Jy/beam
to K for a beam equivalent to the extraction box; sensitivity 
values are given in Table \ref{obstab}. 

The \hcop\ and HCN spectra show red-shifted absorption, in addition to the
stronger blue peak,  as predicted for an infall model. The CS spectrum shows a
stronger blue peak and red-shifted {\bf self}-absorption, but does not
absorb below the continuum.
The \hccn\ spectrum shows no absorption feature.

\section{Models}

\subsection{General Considerations}\label{general}

Unlike earlier observations with lower resolution, these observations
provide essentially a pencil beam toward the center of B335, allowing
``Doppler tomography." The different parts of the line profile probe
different radii with a minimal mixture of off-center emission. 
A velocity field $v(r) \propto r^{-0.5}$ is expected in the innermost
regions for an inside-out collapse model. 
We see no evidence of the outflow at high velocity in the spatial
images, so the line wings probe the inner part of the infall.
The gas at the ambient velocity lies in the outer, static envelope,
while the gas at low offsets from the ambient velocity probes the
outer, but infalling shells.

A line with a high critical or effective density
(e.g., \citealt{1999ARA&A..37..311E, 2015PASP..127..299S})
will have an
optical depth profile that peaks strongly at the radius where the
infall velocity is the absolute value of the offset from the line 
centroid, {\bf if} there is sufficient density at that point in the
cloud to excite molecules to that level. With both density and
temperature increasing inward, these lines will tend to emit
strongly at velocities displaced from the centroid velocity.
If the transition is
optically thin throughout the shell, the resulting line profile
will be double peaked with neither emission nor absorption of the
continuum at the line center. A more opaque line will show the
classic blue asymmetry, in which the blue peak is stronger, with a
minimum of emission at the line center (e.g., the CS \jj76\ line). If
there is sufficient optical depth in the outer layers, 
a line observed toward the continuum peak will show absorption against
the continuum at the velocity centroid from the static envelope
and at redshifted velocities from the outer, infalling shells
(e.g., the HCN and \hcop\ \jj43\ lines).
These features can be seen in models with simple abundance distributions
(\S \ref{infallmods}).

Another possible cause for an absorption dip that is redshifted from the
mean core velocity is an unrelated, redshifted foreground layer, as
was suggested in the case of NGC1333 IRAS4 \citep{2004ApJ...617.1157C}.
The simplicity and isolation of the B335 globule \citep{2007A&A...475..281G}
make this explanation highly unlikely, and no evidence for such a layer
is seen in the many spectra of the source in \citet{2005ApJ...626..919E}
nor in the maps of the region in \citet{2008ApJ...687..389S}.
Furthermore, the lower states of the lines observed here lie much
higher in energy  than those observed in NGC1333 IRAS4, requiring
a warm, dense layer to provide significant optical depth.
The excellent coverage of the uv plane with ALMA, together with the
lack of any foreground molecular gas in B335, largely eliminates the concerns
expressed about the interpretation of the redshifted absorption toward
the NGC1333 region.

To demonstrate further that the absorption feature is confined to the
continuum source, we averaged spectra extracted in circles of
0\farcs5 diameter  located 0\farcs5 
north and south of the center, avoiding the outflow
which goes east and west. These are shown in Figure \ref{modelnopd}.
While red-shifted {\bf self}-absorption can be seen,
the lines no longer show significant absorption below the continuum.
(The \hcop\ line does go slightly below the continuum, but only
at the 2 $\sigma$ level.)

Taken together, these facts indicate that infall has been clearly
detected in the HCN and \hcop\ lines via the IPC signature.
This conclusion is the most important result and 
is independent of the following, more detailed modeling.

\subsection{Simple Models of Line Emission: Pure Infall}\label{infallmods}

Our goal is to see if a previously developed
physical model can account approximately for the observed line profiles.
More complex  models, with asymmetric structures and outflow cavities,
require other constraining data, and are deferred to a later paper.
Because the HCN and \hcop\ line profiles both show red-shifted
absorption, we focus on them here. 

The basic physical model is the inside-out collapse model
\citep{1977ApJ...214..488S},
in which a wave of infall propagates outward at the effective sound speed
(\cseff),
 which is the quadrature sum of thermal and turbulent
contributions.
With \cseff\ constrained by observations, the model
has only one free parameter, the time since collapse began ($t$), 
or equivalently
the infall radius (\rinf) because $\rinf = \cseff t$.
Outside the infall radius (\rinf), the infall velocity is zero; inside
\rinf, the infall velocity increases as $r$ decreases ($v \propto r^{-0.5}$)
and the density distribution is more shallow. The size of \rinf\ is the
best free parameter for comparison to observations.

We use the same modeling apparatus as we used in
\citet{2005ApJ...626..919E}.
Details are given there, but we summarize them here
briefly. A model of inside-out collapse is calculated from the equations in
\citet{1977ApJ...214..488S}, assuming an initial $\tk = 13$ K
\citep{1990ApJ...363..168Z},
 a dust radiative transport code calculates
the dust temperature as a function of radius, and a gas energetics code
calculates the gas temperature as a function of radius including photoelectric
and cosmic ray heating, molecular line cooling, and energy transfer between
dust and gas 
\citep{2004ApJ...614..252Y}.
To this physical model, we add a distribution of molecular abundance for
the species being modeled, calculate the level populations with a
Monte Carlo code for line radiative transfer 
\citep{1995ApJ...448..742C},
and calculate the line profile into a beam of diameter 0\farcs57 to match the
area of the extraction box for the spectra.
To model the source on smaller scales to match the
 the finer resolution of the ALMA observations,
we have doubled the number of shells (to 80) 
in the radiative transport code, extending the inner radius
of the model to 6 AU. 
We have added a continuum source at the center of the model,
using the continuum flux density, converted to radiation temperature.
Once the excitation and radiative transport have been calculated,
we remove a baseline using velocities outside the range of emission
to produce the model spectra 
because the continuum was subtracted from the data. Thus absorption
against the continuum will go below zero in the models, as well as
in the observations.

\begin{figure}
\center
\includegraphics[scale=0.7, angle=0]{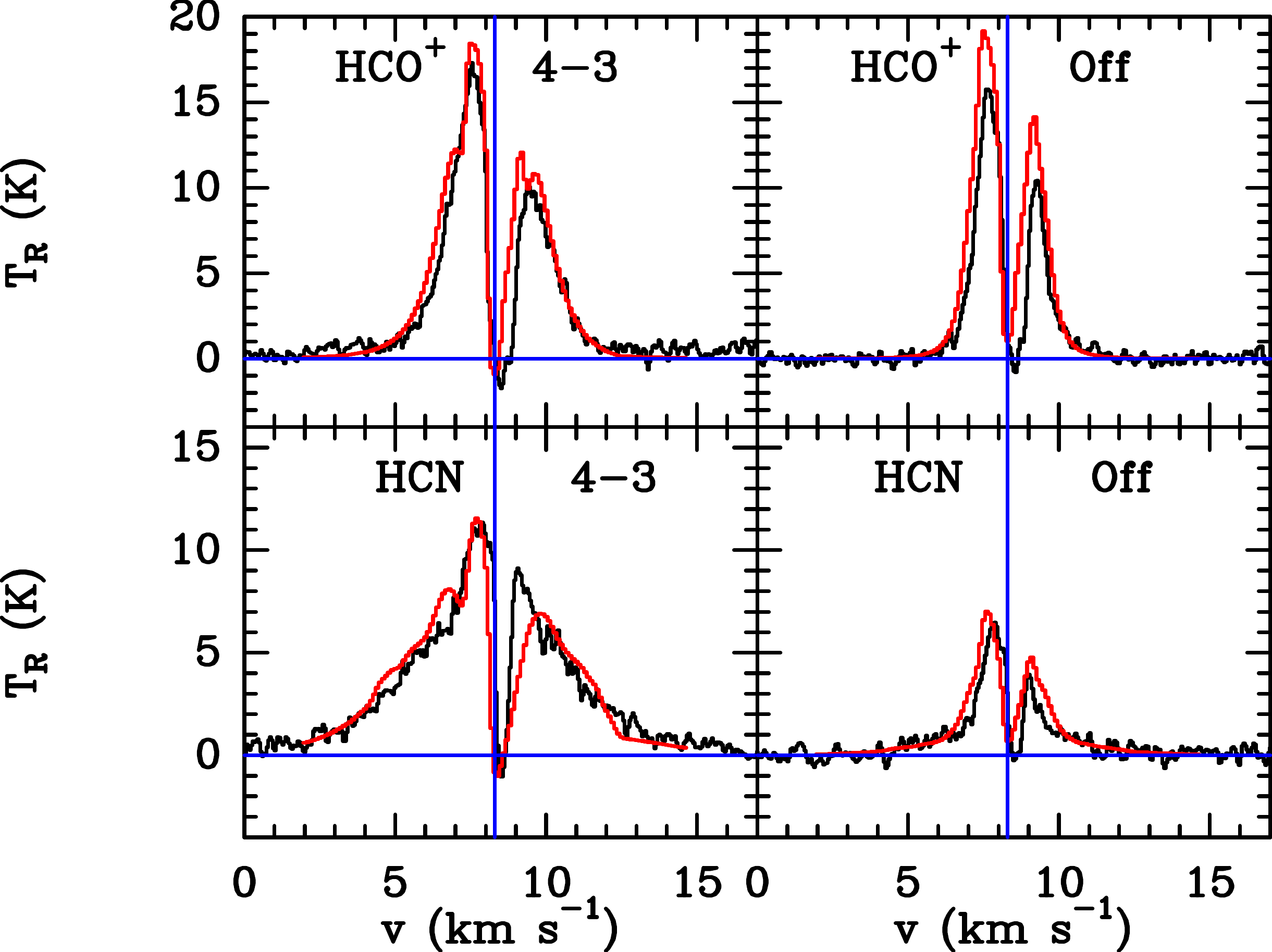}
\caption{Models (red lines) of the lines for \hcop\ and HCN are shown
on top of the observed (black lines) profiles. The observations on the
left are for centered spectra, while those on the right
are the average of two positions, north and south by 0\farcs5
of the peak. The model is for
a spherical inside-out collapse
with $\rinf = 0.012$ pc and abundance profiles in Table
\ref{tableabun}.
The blue horizontal line is at zero and the blue vertical line indicates
the systemic velocity.
}
\label{modelnopd}
\end{figure}

Based on optimizing the fit to many lines observed with
resolutions ranging from 16\arcsec\ to 89\arcsec, 
\citet{2005ApJ...626..919E}
confirmed the best-fit infall radius found by
\citet{1995ApJ...448..742C}, after scaling to
the new distance of 100 pc, of $\rinf = 0.012$ pc. 
Similarly, the luminosity is scaled down to 0.72 \lsun\ at the
new distance.
The ALMA data for HCN and \hcop\ and the predictions of the model 
with $\rinf = 0.012$ pc are shown in Figure \ref{modelnopd}.
The model produces a reasonably good fit to the double-peaked
lines, the absorption dip, and even the line wings.
Models of the off-source spectra (right panels in Figure \ref{modelnopd})
are also reasonably close to the observations; they show self-absorption,
but no absorption against the continuum (a weak continuum, measured from
the observations, of about 10 mJy/beam was included in the model).
These features of both observations and models match the expectations
for line models discussed in \S \ref{general}.
The simple, spherical models do not perfectly match the observations.
More flexible models with flattened envelopes, outflow cavities, etc.
will be explored in future work.

Tests of other infall radii clearly confirm the value of \rinf\
from 
\citet{2005ApJ...626..919E}.
Smaller infall radii fail to predict the width of the
red-shifted absorption and the wings of the lines.
Larger infall radii predict absorption features and line wings
that are wider than observed.
This \rinf\ is also consistent
with the broken power-law model of 
\citet{2013ApJ...765...85K}, who find a break at 0.013 pc,
after adjustment from their assumed 
distance of 150 pc to our assumed distance of 100 pc.

The abundances of the molecules needed tuning on small scales
to match the data.
For the \hcop\ line, we used a step function abundance profile,
in which the abundance drops inside a radius $r_{\rm out}$.
For the HCN line, a drop function 
\citep{2004A&A...416..603J}
was used, in which the abundance drops inside $r_{\rm out}$, but 
rises again within $r_{\rm in}$.
Parameters for both are shown in Table \ref{tableabun}. 
While often used, these abundance models are somewhat
ad hoc and we present them only as a proof of principle that the
absorption feature and the line profile in general can be matched.

The fact that the observed spectra, including both the redshifted
dip and the line wings, can be fitted with a simple model of 
inside-out collapse, with \rinf\ determined from data on much
larger angular scales, albeit with ad hoc abundance variations, gives
us further confidence in our interpretation of the observations as a clear
detection of infall.

\subsection{Updated Model and Uncertainties}

In this section, we update the parameters of the infall model of
\citet{2005ApJ...626..919E} 
to reflect the distance of 100  pc
and discuss some uncertainties in the parameters.
The infall radius of 0.012 pc, together with an effective sound speed 
(combining thermal and turbulent contributions) of 
$\cseff =  0.233$ \kms, implies an age of 5.0\ee4 years since
formation of the central point source, reasonably
consistent with the dynamical timescale for the outflow of 3\ee4 years
(adjusted to 100 pc) from \citet{2008ApJ...687..389S}. 
The mass infall rate would be $\minf = 0.975 c_{\rm s}^3/G = 2.94\ee{-6}$ 
\msunyr, allowing
a central mass (star and disk) of 0.15 \msun\ to  accumulate if the
infall rate has been constant  and mass loss in the wind is neglected.

For comparison,
the mass of gas and dust inside a radius of 25 AU can be estimated from
the continuum emission of 91 mJy, using equation A.31 of
\citet{2008A&A...487..993K}. This quantity is the mass that is not opaque at
846 \micron, so it does not measure mass in a star or the opaque 
part of a disk. 
We use the mass-weighted dust temperature (equation 8) of
\citet{2008A&A...487..993K} of 111 K, based on the model dust temperature
of 67 K at 25 AU. Finally, we use the average opacity at 850 \micron\
determined for B335 itself by 
\citet{2011ApJ...728..143S} 
of 1.48\ee{-2} cm$^2$ per gram of gas for a
gas to dust ratio of 100. The resulting mass is 7.5\ee{-4} \msun, 
much less than the 0.15 \msun\ that has fallen in,
indicating that most of the accumulated central mass is probably in the 
forming star.  
\citet{2015arXiv150904675Y} fit their \cooo\ data with
a smaller central mass, 0.05 \msun; this small a mass is not compatible
with the line wings that we observe.
The limit on other continuum sources in the field translates
to 5\ee{-5} \msun, ensuring that B335 remains single on scales of 100 - 1000
AU.

The mass infall rate can be compared to the rate of accretion onto
the central star (\mdotacc) and to the mass loss rate in the wind 
(\mdotwind) with several further assumptions. 
\begin{equation}
\mdotacc = \frac{\lstar \rstar}{G \mstar \facc} ,
\end{equation}
where \lstar\ is the stellar luminosity, \rstar\ is the stellar radius,
$G$ is the gravitational constant, \mstar\ is the stellar mass, and
\facc\ is the radiative efficiency of accretion (e.g., equation 2 of
\citealt{2014prpl.conf..195D},
where we neglect the photospheric luminosity).
Radiative transfer models of aspherical clouds with outflows
oriented nearly perpendicular to the line of sight
show that observed luminosities can underestimate the
actual luminosity by substantial factors 
(e.g., \citealt{1981A&A....99..289N,2003ApJ...591.1049W, 2012ApJ...753...98O}). 
Preliminary models of B335 with outflow cavities
 suggest that the actual luminosity
is about twice the observed luminosity, so we take $\lstar = 1.5$ \lsun.
If all the mass that has fallen in so far is in the star, $\mstar = 0.15$
\msun. The value of \facc\ is probably less than 0.5, while the value of
\rstar\ depends on the mode of accretion
\citep{2012ApJ...756..118B}. If we take $\facc = 0.5$ and $\rstar = 1.5$
\rsun, we get $\mdotacc = 9.6\ee{-7}$ \msunyr, about one-third of the
infall rate. Like many other young objects, B335 may have a lower 
current accretion rate than infall rate, evidence for
episodic accretion models
\citep{2014prpl.conf..195D}.

The time-averaged mass loss rate in the wind, \mdotwind, can be inferred 
from the force
needed to drive the CO outflow divided by the wind velocity. 
Scaling the force given by
\citet{1992A&A...261..274C}
to a distance of 100 pc yields a value of 1.4\ee{-5} \msunyr\ \kms.
For a wind velocity of 100 \kms, $\mdotwind = 1.4\ee{-7}$ \msunyr,
7 times less than the  accretion rate inferred from the current 
luminosity. A factor of 10 is commonly assumed.

Comparison of all these rates suggests that the current accretion rate is not
dramatically different from its time-averaged value, but that it is 
 less than the infall rate. If mass
is accumulating somewhere besides the star, it cannot be in the optically
thin parts of a disk. A magnetically supported pseudo disk is an interesting
possibility in light of the evidence for magnetic braking
\citep{2015arXiv150904675Y}.

The model used here, spherical inside-out collapse from an isothermal
sphere, is very simple. Because it fits the data well, it makes no
sense to use a more complex models, with more
free parameters, to analyze
these data. However, we discuss briefly the likely effect of
some more complex models. Dense cores are known to have temperature gradients,
with $T$ as low as 7-8 K near the center 
(e.g., \citealt{2001ApJ...557..193E,2007A&A...470..221C, 2013A&A...551A..98L}).
A temperature gradient changes the pre-collapse density structure, but 
the effects are modest \citep{2001ApJ...557..193E}. Also, the infall
wave would propagate more slowly at early times, but this effect is
mitigated by the turbulent contribution to the effective sound speed.
Assuming a constant \tk\ of 8 K, rather than 13 K, would make
the age 6.1\ee4 yr, about 20\% larger.
The initially lower \tk\ would also be counteracted by the heating 
of the interior by the growing stellar luminosity. 
The mass infall rate and central mass are more sensitive to
the sound speed ($\minf \propto \cseff^3$ and $\mstar \propto \cseff^2$, 
respectively).
An initial temperature of 8 K would yield $\minf = 1.6\ee{-6}$ \msunyr\
and $\mstar = 0.097$ \msun. Interestingly, experiments with initial
temperatures even as low as 10 K produced worse fits to the HCN line.

Rotation will modify the evolution, but other studies
have shown that B335 has very little rotation even at small
scales, probably because of magnetic braking 
\citep{2015arXiv150904675Y}.
The most important complication is the existence of the outflow,
which can cause an underestimate of the luminosity and hence temperature
of the gas.
More direct effects, such as confusing the infall signature
with outflow emission, seem to be unimportant at the resolution of
these observations.

\section{Summary}

The \hcop\ and HCN lines observed with 0\farcs5 resolution by ALMA
show red-shifted absorption, as expected for a model of infalling gas.
A model of infall from a singular isothermal sphere developed previously
from data with much lower spatial resolution can fit the new data, when
updated with the much closer distance and with some tuning of abundances.
B335 is clearly a special source in that it is isolated and close to us.
However, it fits in, toward the low luminosity end, with a group of
very embedded sources found recently; from the fraction of all
protostars in this phase
\citet{2013ApJ...767...36S} estimated a lifetime of 2.5\ee4 yr, comparable
to what we find for B335 with a completely different method.
Further studies of this source can shed light on the larger class
of very young protostars.

\medskip

We are grateful to A. Stutz and H-W. Yen for comments on a previous version.
This paper makes use of the following ALMA data: ADS/JAO.ALMA\#2012.1.00346.S. 
ALMA is a partnership of ESO (representing its member states), NSF (USA) and 
NINS (Japan), together with NRC (Canada) and NSC and ASIAA (Taiwan), in 
cooperation with the Republic of Chile. The Joint ALMA Observatory is 
operated by ESO, AUI/NRAO and NAOJ. This work was supported by NSF
grant AST-1109116 to the University of Texas at Austin.
The research of JKJ is supported by a Lundbeck Foundation Junior Group 
Leader Fellowship. J.-E. Lee was supported by the Basic Science Research 
Program through the National Research Foundation of Korea (NRF) 
(grant No. NRF-2015R1A2A2A01004769) 
and the Korea Astronomy and Space Science Institute under the R\&D 
program (Project No. 2015-1-320-18) supervised by the Ministry of Science, 
ICT and Future Planning.

\bibliographystyle{apj}


\begin{table}[h]
\caption{ALMA Observations} \label{obstab}
\vspace {3mm}
\begin{tabular}{l r r r r r  }
\tableline
\tableline
Transition &  Frequency  & $\delta v$ & $E_{u}$ & $A_{ul}$ 
& Sensitivity    \cr
           &  (GHz)      & (\kms)               & (K)     & (s$^{-1}$) 
& (K/(Jy/bm)$^{-1}$) \cr
\tableline
CS \jj76\      & 342.882860  & 0.107 & 65.8    & 8.4\ee{-4}  & 32.0  \cr
\hccn\ \jj43\  & 345.339760  & 0.106 & 41.4    & 1.90\ee{-3} & 31.6 \cr
HCN \jj43\     & 354.505470  & 0.103 & 42.5    & 2.05\ee{-3} & 30.0 \cr
\hcop\ \jj43\  & 356.734240  & 0.103 & 42.8    & 3.57\ee{-3} & 29.6  \cr
\tableline
\end{tabular}
\end{table}

\begin{table}[h]
\caption{Molecular Abundances} \label{tableabun}
\vspace {3mm}
\begin{tabular}{l r r r r  }
\tableline
\tableline
Species &  $X$ & $X_{\rm drop}$ & $r_{\rm out}$ & $r_{\rm in}$    \cr
        &      &                &  (pc)         &   (pc)          \cr
\tableline
\hcop\         &3.0\ee{-9} & 1.5\ee{-10} & 1.2\ee{-3}  & \ldots  \cr
HCN            &1.5\ee{-9}  & 7.5\ee{-11}& 1.2\ee{-3}  & 1\ee{-4}   \cr
\tableline
\end{tabular}
\end{table}

\end{document}